\documentstyle[twocolumn,prl,aps]{revtex}            
\newcommand{\Temp}{\Theta}

\begin{document}
\draft

\twocolumn[\hsize\textwidth\columnwidth\hsize\csname 
@twocolumnfalse\endcsname                            

\title{Solidification in a channel}

\author{Mohsen Sabouri-Ghomi$^{1}$, Nikolas Provatas$^{1,2}$, and
Martin Grant$^{1}$}

\address{$^1$Physics Department and Centre for the Physics of Materials,
Rutherford Building, McGill University, 3600 rue University, Montr{\'e}al,
Qu\'ebec, Canada H3A 2T8}

\address{$^2$Pulp and Paper Research Institute of Canada, 570 boulevard
St.-Jean, Pointe-Claire, Qu\'ebec, Canada H9R 3J9}

\date{\today}

\maketitle

\begin{abstract}

We simulate solidification in a narrow channel through the use of
a phase-field model with an adaptive grid.  In different regimes,
we find that the solid can grow in fingerlike steady-state shapes, or
become unstable, exhibiting unsteady growth.  At low melt undercoolings,
we find good agreement between our results, theoretical predictions,
and experiment.  For high undercoolings, we report evidence for a new
stable steady-state finger shape which exists in experimentally accessible
ranges for typical materials.

\end{abstract}

 


\vskip2pc]                                       

The classic example of pattern selection in driven systems is the
shape of a dendritic crystal's tip, which is growing into a supercooled
melt, mediated by the diffusion of the latent heat of solidification
\cite{LANGER,REVIEWS}.  Simple theories, which neglect the effect
of surface tension $\kappa$, predict that only the product of the
velocity $V$ and dendrite tip radius $\rho$ is determined for a given
undercooling.  However, the modern theory of microscopic solvability,
which incorporates the surface tension, predicts the unique operating
condition for a steady-state tip, provided that $\kappa$ is
anisotropic\cite{REVIEWS}.
This work generalized earlier work on local models of solidification
\cite{GEOMETRIC_MODEL,BOUNDARY_LAYER_MODEL} and, in particular, work
on the closely related problem of viscous fingering in Hele-Shaw cells
\cite{VISCOUS}.  The latter phenomena corresponds to the fingering
instability which occurs when a relatively inviscid fluid, such as
water, is forced by a pressure gradient to invade a more viscous
fluid, such as oil.  It can be shown that, if this process takes
place between two glass plates which form a long channel  of width $L$
--- called a Hele-Shaw cell --- a long steady-state finger forms and
grows \cite{saffman-taylor}.  As in the case for dendritic tips, in
the absence of surface tension, theory predicts fingers of any width.
The theory of microscopic solvability predicts a unique finger width,
e.g., in units of the channel width, the finger width approaches $\lambda
= 1/2$ in the limit of $\kappa \rightarrow 0$.  Mathematically, both the
dendrite tip and the viscous finger are described by similar equations,
namely, diffusion or the Laplace problem in the bulk phases, subject to
the boundary conditions of the moving interface, and the driving force
at infinity far from the interface.  In the limit in which the diffusion
length $l_D$ becomes very large compared to the channel width, where $l_D
= D/V$ and $D$ is the diffusion coefficient (that is, if the P\'eclet
number $P = L/l_D$ is small), the two problems become almost identical.
This close relationship has led to many beautiful insights into both
phenomena \cite{kkl86,libchaber90,SCRATCH}.  For example, the ``free''
viscous finger generically shows unsteady tip splitting behavior, which
can be attributed to the fluid's isotropic surface tension.  However,
creating an effective anisotropic surface tension, by scratching one of
the glass plates, stabilizes the tip, giving rise to dendritic patterns
\cite{SCRATCH}.

Here we report the results of a theoretical investigation into the
alternate case: solidification constrained to grow in two dimensions
within a long channel.  For small $P$, this is closely related to
viscous fingering with anisotropic surface tension.  We compare our
numerical results for a phase-field model of solidification, with no
adjustable parameters, to theory \cite{kkl86} and experiment 
\cite{libchaber90}, finding good agreement,
thereby providing the first quantitative test of solvability theory
in this context.  For larger $P$, the conservation law for the latent
heat plays an important role.  Indeed, for large undercoolings, we find
evidence for a new steady-state finger, which we relate to local models
\cite{GEOMETRIC_MODEL,BOUNDARY_LAYER_MODEL} of dendritic growth.

Our approach takes advantage of recent advances in computer power, as 
well as algorithmic development.  We use a two-dimensional phase-field
model of solidification,
\cite{kobayashi,EARLY,wheeler,elder,wang,KARMA,nick98,nick2000,SHARP}
where a scalar field $\phi(\vec x, t)$ determines whether the phase
at field point $\vec x$ and time $t$ is solid ($\phi=+1$) or liquid
($\phi=-1$).   A useful approach introduced in Ref.\ \cite{KARMA}
allows for the elimination of kinetic undercooling.  As well, we
use finite-element adaptive-grid refinement techniques, introduced
by one of us and collaborators \cite{nick98}, wherein a fine mesh is
used near the rapidly varying solid-melt interface, and a coarse mesh
is used in the bulk phases where quantities vary slowly with space and 
time.  Let the dimensionless temperature be
$\Temp =c_{p}(T-T_{M})/\ell$,
where $c_{p}$ is the specific heat at constant pressure, $T_{M}$ is the
melting temperature, and $\ell$ is the latent heat of the fusion.
Then the equations of motion are \cite{nick98,nick2000} 
\[
\frac{\partial \Temp}{dt} = D \nabla^2 \Temp + \frac{1}{2} \frac{\partial
\phi}{\partial t},
\]
and,
\begin{eqnarray}
A^2(\vec{n}) \frac{\partial \phi}{dt} &=& \vec \nabla \cdot
[A^2(\vec{n}) \vec \nabla \phi ]  +
[\phi - \gamma \Temp (1-\phi^2)][1- \phi^2] \nonumber \\
&+&
\frac{1}{2}\vec \nabla \cdot \left[ |\nabla \phi|^2  
\frac {\partial A^2(\vec{n})}{\partial (\vec \nabla \phi)} \right], 
\nonumber
\end{eqnarray}
where $D={\cal {D}} \tau_0/W_0^2$, ${\cal {D}}$ is the thermal diffusivity.
All quantities were measured in units of the microscopic time and
length scales, $\tau_0$ and $W_0$.  The parameter $\gamma$ is chosen
to be $1.5957 D$ to eliminate kinetic undercooling \cite{KARMA}, and
thereby relate  $\tau_0$ and $W_0$ to experimental values.
Anisotropy enters through $A(\vec n)$, where $\vec{n} = \vec \nabla
\phi/|\nabla \phi|$, is the unit vector normal to the contours of $\phi$:
$A(\vec{n}) = [1 - 3 \epsilon] [ 1 + \frac{4 \epsilon}{1 - 3 \epsilon}
((n_x)^4 + (n_y)^4 )]$, where $\epsilon$ is a constant  \cite{nick98}.
We used $\epsilon = 0.05$ \cite{libchaber90,WEASELTALK}. The width
of the interface is $W(\vec{n})=W_0 A(\vec{n})$ and the characteristic
time $\tau(\vec{n})=\tau_0 A^2(\vec{n})$ \cite{KARMA}.  Using $d_{o}
= 0.5539/D$, the boundary condition at the interface \cite{SHARP} is
$\Temp_{int} = -d(\vec n) K$, where $K$ is the interface curvature, and
$d(\vec n) = d_0[A(\vec n)+\partial^2 A/\partial({\cos^{-1} n_x})^{2}]$
is the capillary length.  The simulations were performed using Neumann
boundary conditions: $\vec{n} \cdot \vec \nabla \phi = 0 $ and $
\vec{n} \cdot \vec \nabla \Temp = 0$, so that heat cannot leak out of
the channel.  Solidification is initiated by a small half disk of radius
$R_0$ centered at the bottom of the channel.  The interface is initially
set to its equilibrium value $\phi(\vec x) = -\tanh (x - R_0)/\sqrt2$.
The initial temperature is $\Temp = 0$ in the solid phase, which decays
exponentially to its asymptotic value $\Temp = \Delta$ at $x\rightarrow
\infty$ in the far field.  The time step was chosen to be 0.016, and
the minimum grid size for the adaptive grid was 0.78 and 0.39, for the
low and high $P$ regimes, respectively. 

With this approach, we can make a quantitative comparison to experiment
as well as theory, since as $P\rightarrow 0$ these equations
reduce to those of the anisotropic viscous finger.  Motivated by
the experiment of Molho, Simon, and Libchaber \cite{libchaber90},
the first set of simulations were done in a wide channel $L/d_0 =
121,000$ with rectangular cross section of $1600W_0$ by $102400W_0$
(319.3 $\mu$m $\times$ 2 cm), similar to the experimental capillary
cells \cite{libchaber90} ($200 \mu$m $\times$ 6 cm).  To examine the low
P\'eclet regime, the simulations were done at low undercoolings $\Delta$
= 0.01, 0.05, 0.1, 0.125, 0.15, with $D=40$ and $R_0=30$.  After a short
period of acceleration, tip velocity decreases to an asymptotic value,
following approximately an inverse square root of time.  To estimate
this asymptotic tip velocity and width of the fingers, we extrapolated
our transient results using a scaling ansatz, as in Ref.~\cite{nick2000},
and by fitting to a high-order polynomial.  The scaling ansatz is shown
as an inset in Fig.~1; we scaled the width of the finger with $t^\alpha$,
and the length of the finger with $t^\beta$, obtaining very good data
collapse for the entire shape of the finger for different undercoolings
\cite{ALPHABETA}.

This method and the polynomial fits gave consistent results for $\lambda$,
which are shown in Fig.~1, where $V$ appears through the dimensionless
surface tension $\kappa = \frac{D d_{0}\pi^{2}}{L^{2} V (1-\lambda)^2}$.
We found that for these channel dimensions the dimensionless finger
width, $\lambda$, is smaller than 0.5, and decreases with increasing
velocity, as predicted in the solvability theory of Kessler, Koplik,
and Levine \cite{kkl86}.  Also shown are the results of the experiment,
and the solvability solution for the anisotropic viscous finger
\cite{kkl86,dorsey,eugenia-hong}.  With no adjustable parameters,
our phase-field model results are consistent with the experimental
observations \cite{libchaber90,WEASELTALK} and compare
very well to the theoretical prediction \cite{kkl86,dorsey,eugenia-hong}.
Our results provide the first quantitative test of that prediction of
solvability theory for anisotropic viscous fingers, confirming those
predictions.

This correspondence between solidification in a channel and the
anisotropic viscous finger, which we have exploited above, is restricted
to small P\'eclet numbers, and to a transient time regime which diverges
as $P\rightarrow 0$.  For later times, and larger $P$, the situation
is different: from thermodynamics, the only asymptotic solution for
an arbitrarily long solid finger within a liquid channel is $\lambda
=\Delta$, since $\Delta$ is the fractional amount of solid in equilibrium
\cite{kkl86,pelce-euro}.  For long times, far beyond those investigated
above, it is expected that the finger shape obtained above is unstable
due to latent heat piling up in front of the finger.  For example,
Pelce \cite{pelce-euro} has pointed out that the finger is unstable
to a tip widening instability unless $d\lambda /dV > 0$:  If $\lambda$
is a decreasing function of $V$, then a perturbation giving rise to a
wider finger causes the finger to slow down, leaving more heat in the
vicinity, causing it to slow down more, and flatten out until the finger
reaches the size of the channel.

To investigate this regime, we simulated larger P\'eclet numbers in
a narrow channel $L/d_0 = 355$, of dimension $100W_0$ by $1600W_0$.
The simulations were performed at $\Delta = 0.25 - 0.95$, with  $D=2$
and $R_0 =4$.  Fig.~2 shows the dependence of $V$ on time.  There is a
marked qualitative change at $\Delta \approx 0.5$.  For  $\Delta < 0.5$,
a tip widening instability takes place.  If we estimate an effective
$V$ and $\lambda$ from a transient regime before this instability
takes place, we find that $\lambda$ is a decreasing function of $V$,
as would be expected: the dependence of $\lambda$ on undercooling for
these fingers is shown by the open circles in Fig.~3; inset (a) of Fig.\
3 shows one of these fingers while the instability is taking place.

In contrast, for $\Delta > 0.5$, in the presence of anisotropic
surface tension, the fingers rapidly achieve a well-defined constant
velocity (closed circles in Fig.~2) and shape.  Fits to $V$ give $P =
5.4 - 49$.  In Fig.~3, we show that their finger widths obey $\lambda
=\Delta$; inset (b) shows one of these fingers.  Hence these fingers
satisfy thermodynamic and stability requirements, since $\Delta \propto
V$.  This new class of {\it stable\/} fingers fingers becomes unstable
to tip splitting if the surface tension is isotropic.  We do not have
an explanation for the apparent stability criterion $\lambda \ge 0.5$,
although it is interesting to note the reappearance of this bound
for isotropic viscous fingers in the new context of rapidly growing
anisotropic solid fingers.  Nevertheless, for high undercoolings close
to unity, we can suggest an origin for these new fingers.

For high undercoolings, the thermal boundary layer around the solidifying
finger becomes small.  As shown by Ben-Jacob, Goldenfeld, Langer, and
Sch\"on \cite{BOUNDARY_LAYER_MODEL}, in the limit $\Delta \rightarrow
1$ the interface velocity and curvature are determined locally,
as in, e.g., the geometric \cite{GEOMETRIC_MODEL} and boundary-layer
\cite{BOUNDARY_LAYER_MODEL} models of solidification.  In such models,
the steady-state shape is determined by the curvature $K$ as a function
of only the normal velocity in the limit of vanishing surface tension.
That is, for small $K$ and $n_x$, $K = f(n_x) = a n_x + b n_x^3 +
\cdots$.  For the free boundary-layer model \cite{BOUNDARY_LAYER_MODEL},
the coefficient $a$ is exactly zero, and $b \propto (1-\Delta)$, giving
rise to a parabolic shape.  However, it is easy to show that a nonzero
$a$ is required for a finger of finite width; here the value of $a$
is determined by the thermodynamic requirement $\lambda = \Delta$.
This gives rise to the new steady-state solution: 
\[ 
x = -B(\Delta/\pi)\ln \cos (\pi y/\Delta) 
\]
where $1/B= \sqrt{1 + (b\Delta/2\pi)^2} - (b\Delta/2\pi)$.  This form
gives good one-parameter fits to the fingers as shown in Fig.~4 (and
fits which are significantly better than those from the Saffman-Taylor
\cite{saffman-taylor} shape $b\equiv 0$).  Since the form is obtained from
an expansion in $n_x$, it is a better fit in the tails that at the tips.
Nevertheless, the fitted dependence of $b$ on $\Delta$ is reasonably
well described by $b \propto (1-\Delta)$, for $\Delta \rightarrow 1$
as expected.

To conclude, we have investigated solidification in a channel in
low and high undercooling regimes.  Our results for low undercooling
are consistent with earlier experimental work, and compare well to
theory, providing the first quantitative test of solvability theory
for anisotropic viscous fingers.  For large undercoolings, we find new
phenomena, a solid finger which satisfies stability and thermodynamic
criterion.  We further argue for an analytic form of the shape, based on
local models of solidification, which fits our results from numerical
simulation.  We note that this new ``finger'' can be studied further
experimentally and theoretically.

This work was supported by the Natural Sciences and Engineering Research
Council of Canada, {\it le Fonds pour la Formation de Chercheurs et l'Aide
\`a la Recherche du Qu\'ebec\/}.  We thank Hong Guo, Fran\c cois Drolet,
Mikko Haataja, and Martin Dub\'e for useful discussions.

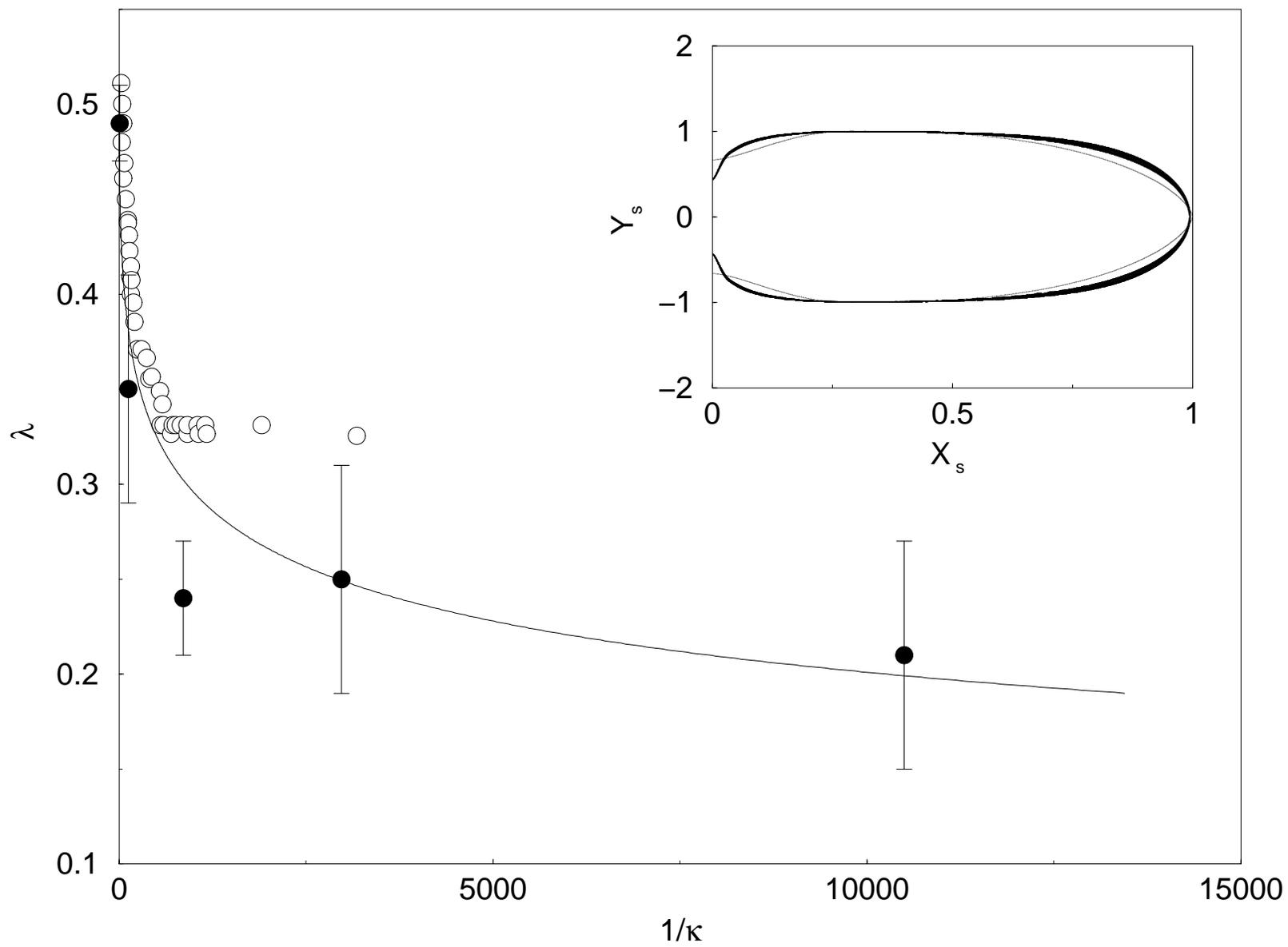
\begin{figure} 
\caption{
Finger width $\lambda$ vs.\ velocity $V$ for small P\'eclet numbers.
Solvability theory [7,21,22] (solid line, $\epsilon = 0.05$),
experimental results [8] (empty circles),
and present simulations (filled circles, $\epsilon=0.05$). 
Inset: scaling collapse used to estimate $V$
for $\Delta = 0.05$ (thin line),
and $\Delta = 0.15$ (thick lines). 
$X_{s}$ is scaled length of finger, $Y_s$ is scaled
width.  For $\Delta = 0.05$ all profiles collapse. 
For $\Delta = 0.15$, seven profiles 
are shown, which collapse quickly for late times.
} 
\label{fig1} 
\end{figure}

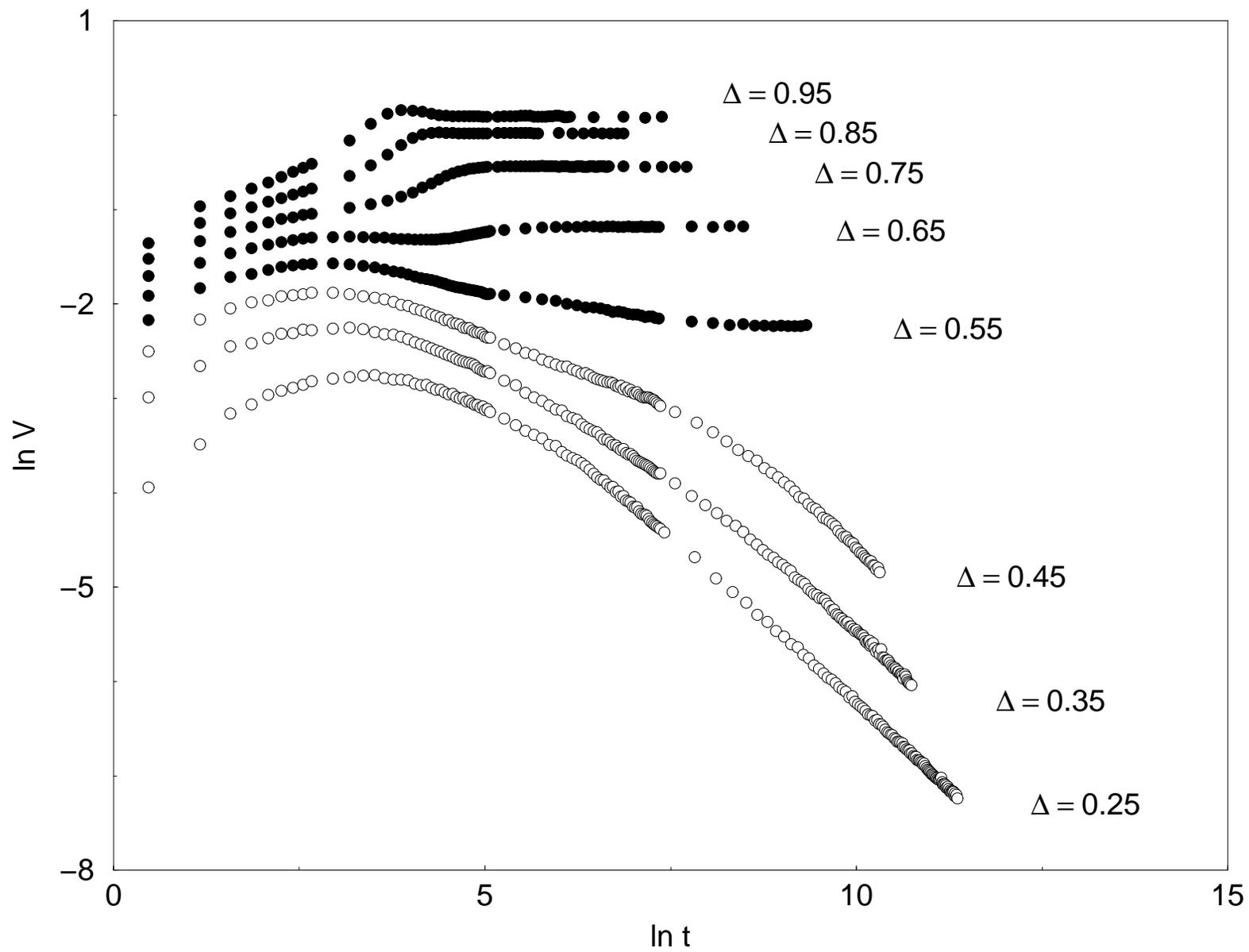
\begin{figure}
\caption{
Velocity $V$ vs.\ time for various undercoolings $\Delta$, for
larger P\'eclet numbers.  Fingers with tip-widening instability
$\Delta  < 0.5$ (open circles).
Note qualitative change in velocities for stable fingers 
$\Delta > 0.5$ (filled circles).
}
\label{fig2}
\end{figure}

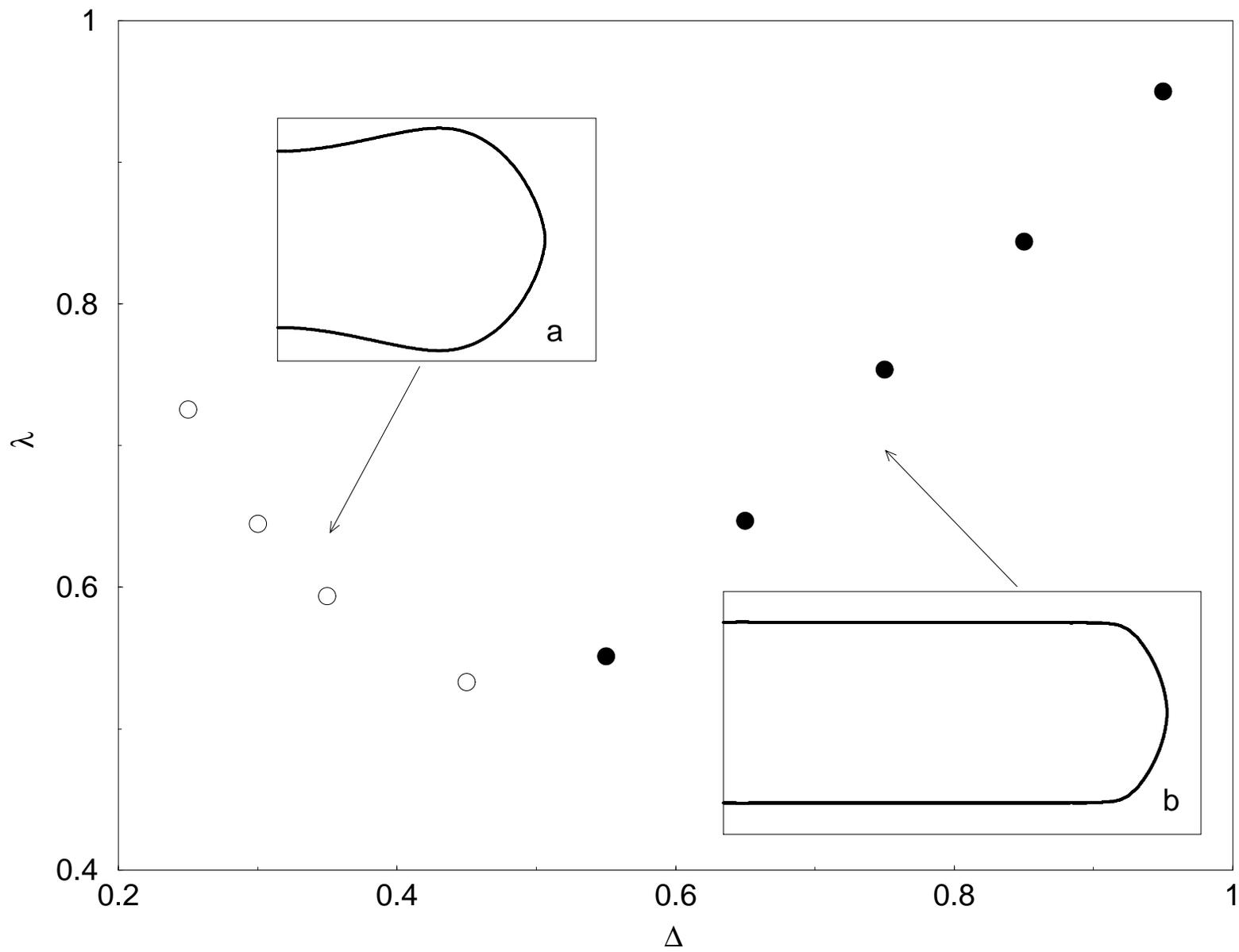
\begin{figure} 
\caption{
Finger width $\lambda$ vs.\ undercooling $\Delta$ for larger P\'eclet
numbers.  Fingers with tip-widening instability (open circles, inset (a)).
Stable fingers (filled circles, inset (b)).
} \label{fig3} 
\end{figure}

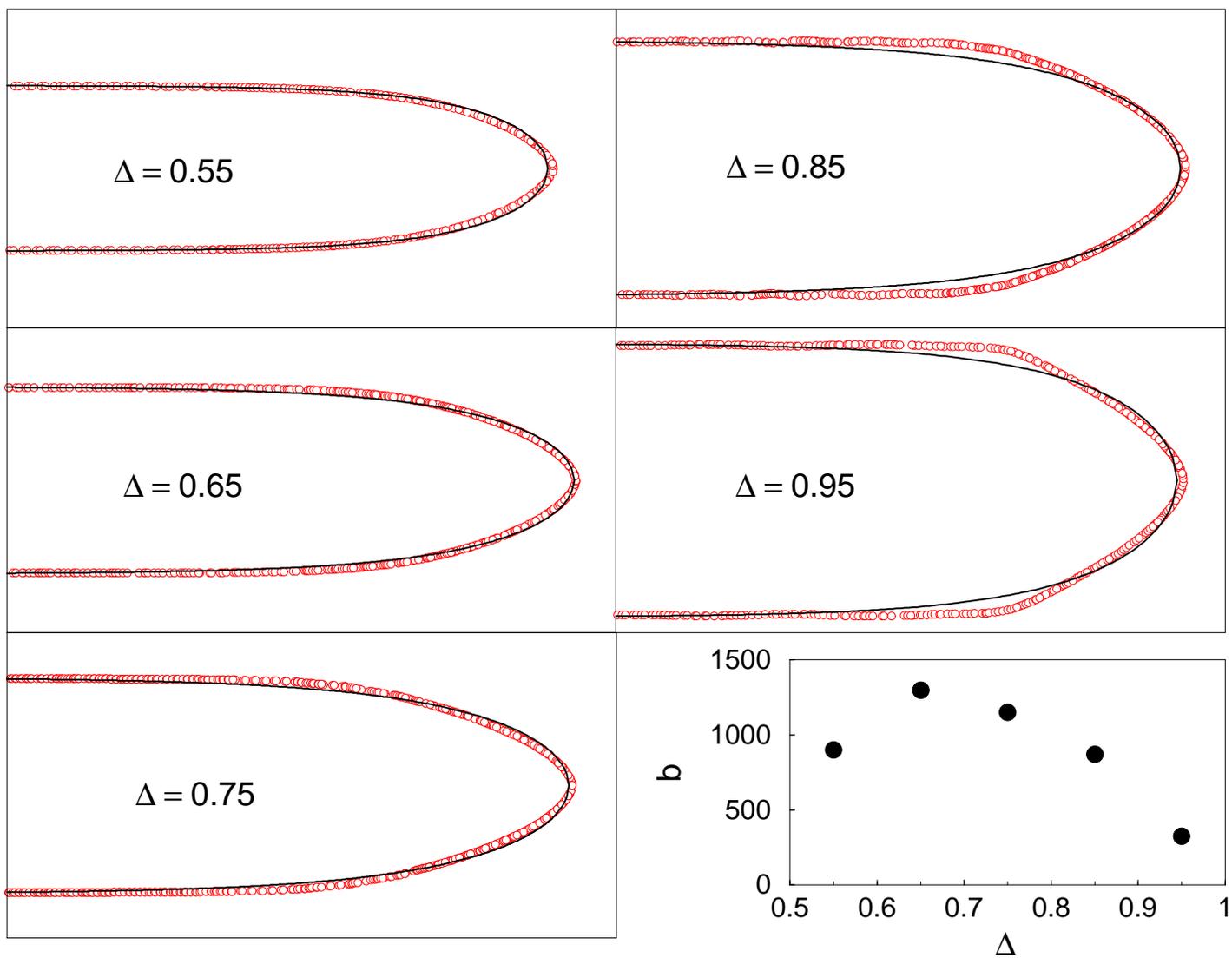
\begin{figure} 
\caption{
Numerical (open circles), and analytic (lines) results for stable finger
tip profiles, at undercoolings $\Delta = 0.55$ -- 0.95.  Analytic fits,
from an expansion valid in the tail, involve one parameter which should
obey $b \propto (1 -\Delta)$ for $\Delta \rightarrow 1$.  $b$ vs.\
$\Delta$ (bottom inset) shows reasonable agreement.  
} 
\label{fig4} 
\end{figure}

\end{document}